\documentclass[aps,preprint]{revtex4}
\usepackage{amssymb}

\usepackage{amsmath}
\usepackage[dvips]{graphicx}

\input{tcilatex}

\begin{document}

\title{Frozen light in periodic stacks of anisotropic layers}
\author{J. Ballato}
\affiliation{Center for Optical Materials Science and Engineering Technologies (COMSET),
School of Materials Science and Engineering. Clemson University, Clemson, SC 29634-0971}
\author{A. Ballato}
\affiliation{US Army Communications - Electronics Research, Development, and Engineering
Center (CERDEC). Fort Monmouth, NJ 07703-5201}
\author{A. Figotin and I. Vitebskiy }
\affiliation{Department of Mathematics, University of California at Irvine, CA 92697}

\begin{abstract}
We consider a plane electromagnetic wave incident on a periodic stack of
dielectric layers. One of the alternating layers has an anisotropic refractive
index with an oblique orientation of the principal axis relative to the normal
to the layers. It was shown recently that an obliquely incident light, upon
entering such a periodic stack, can be converted into an abnormal
\emph{axially frozen mode }with drastically enhanced amplitude and zero normal
component of the group velocity. The stack reflectivity at this point can be
very low, implying nearly total conversion of the incident light into the
frozen mode with huge energy density, compared to that of the incident light.
Supposedly, the frozen mode regime requires strong birefringence in the
anisotropic layers -- by an order of magnitude stronger than that available in
common anisotropic dielectric materials. In this paper we show how to overcome
the above problem by exploiting higher frequency bands of the photonic
spectrum. We prove that a robust frozen mode regime at optical wavelengths can
be realized in stacks composed of common anisotropic materials, such as
$YVO_{4}$, $LiNb$, $CaCO_{3}$, and the like.

\end{abstract}
\maketitle

\section{Introduction}

\begin{figure}[tbph]
\includegraphics[scale=0.9, viewport= 0 0 400 400, clip]{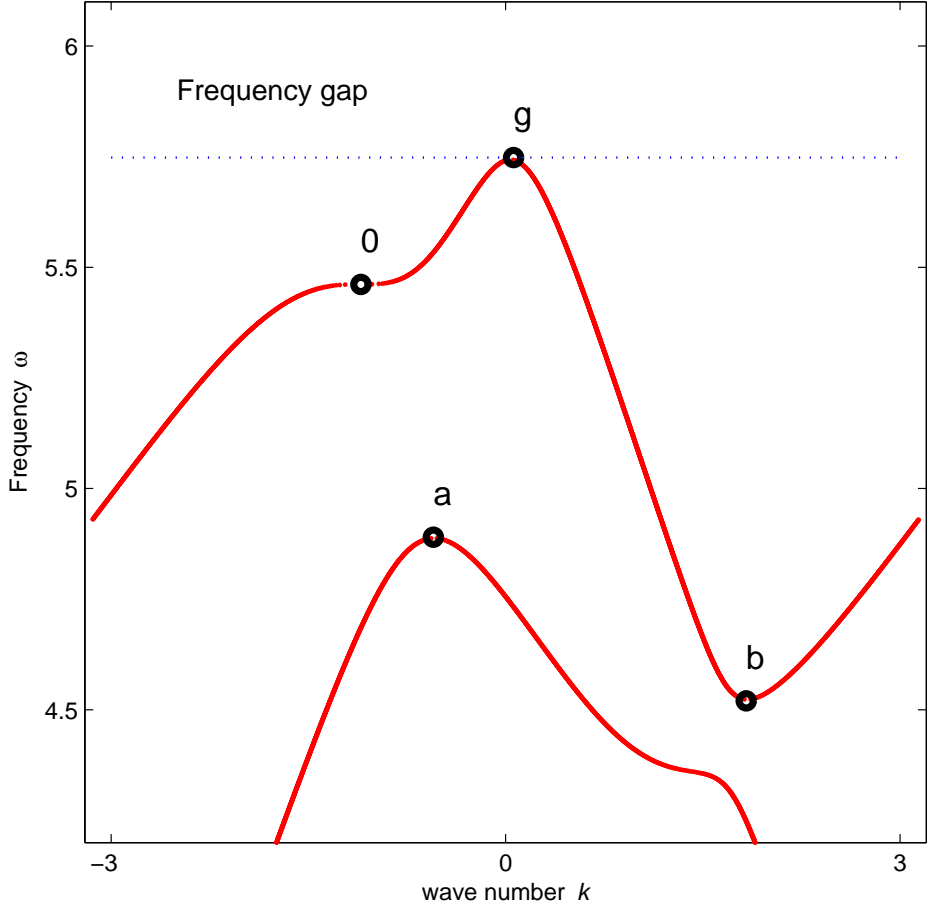}
\caption{An example of electromagnetic dispersion relation $\omega(k)$ with
various stationary points: (i) extreme points $a$ and $b$ of the respective
spectral branches, (ii) a photonic band edge $g$, (iii) a stationary
inflection point $0$. Each stationary point is associated with slow light. The
wave number $k$ and the frequency $\omega$ are given in units of $1/L$ and
$c/L$, respectively. $L$ is the primitive translation of the periodic array.}
\label{DR_PRE03}
\end{figure}

In photonic crystals, the speed of light is defined as the wave group
velocity, $\vec{u}$,%
\begin{equation}
\vec{u}=\partial\omega/\partial\vec{k}, \label{u}%
\end{equation}
where $\vec{k}$ is the Bloch wave vector and $\omega=\omega\left(  \vec
{k}\right)  $ is the respective frequency. At certain frequencies, the
dispersion relation $\omega\left(  \vec{k}\right)  $ of a photonic crystal
develops stationary points%
\begin{equation}
\partial\omega/\partial\vec{k}=0, \label{SP}%
\end{equation}
in the vicinity of which the group velocity vanishes. Zero group velocity
usually implies that the respective Bloch eigenmode does not transfer
electromagnetic energy. Indeed, with certain reservations, the energy flux
$\vec{S}$ of a Bloch mode is%
\begin{equation}
\vec{S}=W\vec{u} \label{S=Wu}%
\end{equation}
where $W$ is the electromagnetic energy density, associated with this mode. If
$W$ is limited, then the group velocity $\vec{u}$ and the energy flux $\vec
{S}$ vanish simultaneously at any stationary point (\ref{SP}) of the
dispersion relation. Such modes are commonly referred to as slow modes, or
slow light. Examples of different stationary points (\ref{SP}) are shown in
Fig. \ref{DR_PRE03}., where each of the respective frequencies $\omega_{a}$,
$\omega_{b}$, $\omega_{g}$ and $\omega_{0}$ is associated with slow light.%

\begin{figure}[tbph]
\includegraphics[scale=0.9, viewport= 0 0 500 400, clip]{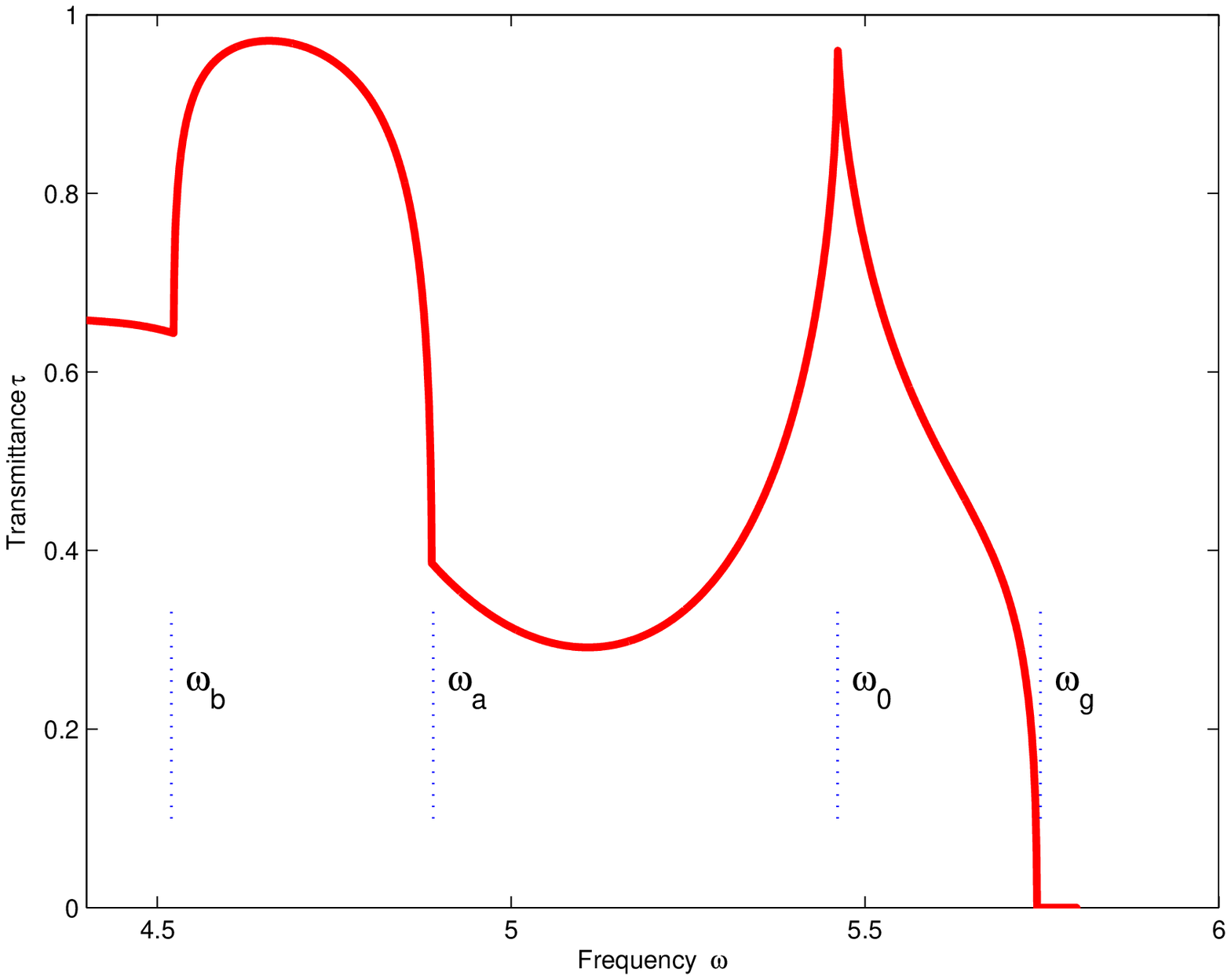}
\caption{Transmittance $\tau$ as a function of incident light frequency
$\omega$ for the semi-infinite photonic slab with the dispersion relation
presented in Fig. 1. The characteristic frequencies $\omega_{a}$, $\omega_{b}%
$, $\omega_{0}$, and $\omega_{g}$ are associated with the respective
stationary points in Fig. 1. At $\omega\geq\omega_{g}$ (within the photonic
band gap) the incident light is totally reflected by the slab.}
\label{tE_PRE03}
\end{figure}

A common problem with slow modes is that most of them cannot be excited in
semi-infinite photonic crystal by incident light. Indeed, consider plane
monochromatic wave incident on a semi-infinite photonic crystal with the
electromagnetic dispersion relation shown in Fig. \ref{DR_PRE03}. If the
frequency $\omega$ is close to the band edge frequency $\omega_{g}$ in Fig.
\ref{DR_PRE03}, then the incident wave will be totally reflected back into
space, as seen in Fig. \ref{tE_PRE03}.%

In another case, where the incident wave frequency is close to the
characteristic value $\omega_{a}$ or $\omega_{b}$ in Fig. \ref{DR_PRE03}, some
portion of the incident wave will be transmitted in the photonic crystal, but
none in the form of the slow mode corresponding to the respective stationary
point. This means, for example, that at the frequency $\omega_{a}$, the
transmitted light is a Bloch wave with the finite group velocity and the wave
number different from that corresponding to the point $a$ in Fig.
\ref{DR_PRE03}.

Let us turn now to the stationary inflection point $0$ of the dispersion
relation in Fig. \ref{DR_PRE03}, where both the first and the second
derivatives of the frequency $\omega$ with respect to $k$ vanish, while the
third derivative is finite%
\begin{equation}
\text{at }\omega=\omega_{0}\text{ \ and \ }k=k_{0}\text{ : \ }\frac
{\partial\omega}{\partial k}=0;\;\frac{\partial^{2}\omega}{\partial k^{2}%
}=0;\;\frac{\partial^{3}\omega}{\partial k^{3}}\neq0. \label{SIP}%
\end{equation}
The frequency $\omega_{0}$ of stationary inflection point is associated with
the so-called \emph{frozen mode regime} \cite{PRE01,PRB03,PRE03}. In such a
case, the incident plane wave can be transmitted into the photonic crystal
with little reflection, as seen in Fig. \ref{tE_PRE03}. Having entered the
photonic slab, the light is 100\% converted into the slow mode with
infinitesimal group velocity and drastically enhanced amplitude. Under the
frozen mode regime, vanishingly small group velocity $\vec{u}$ in Eq.
(\ref{S=Wu}) is offset by a huge value of the energy density $W$. As the
result, the energy flux (\ref{S=Wu}) associated with the frozen mode remains
finite and comparable with that of the incident wave. In the vicinity of the
frozen mode frequency $\omega_{0}$, the electromagnetic energy density $W$
associated with the slow (frozen) mode displays a resonance-like behavior
\begin{equation}
W\approx\frac{2\tau S_{I}}{6^{2/3}}\left(  \omega_{0}^{\prime\prime\prime
}\right)  ^{-1/3}\left(  \omega-\omega_{0}\right)  ^{-2/3}, \label{W(SIP)}%
\end{equation}
where $S_{I}$ is the fixed energy flux of the incident wave, $\tau$ is the
portion of the incident light transmitted into the semi-infinite photonic
crystal, and%
\[
\omega_{0}^{\prime\prime\prime}=\left(  \frac{\partial^{3}\omega}{\partial
k^{3}}\right)  _{k=k_{0}}.
\]
Remarkably, the transmittance $\tau$ at $\omega\approx\omega_{0}$ remains
finite and may even be close to unity, as shown in Fig. \ref{tE_PRE03}. The
latter implies that a significant portion of the incident light is converted
into the frozen mode with nearly zero group velocity and huge amplitude,
compared to that of the incident wave. In reality, the electromagnetic energy
density $W$ of the frozen mode will be limited by such factors as absorption,
nonlinear effects, imperfection of the periodic dielectric array, deviation of
the incident radiation from a perfect plane monochromatic wave, finiteness of
the photonic slab dimensions, etc. Still, with all these limitations in place,
the frozen mode regime can be very attractive for various practical applications.

From now on we restrict ourselves to the case of lossless periodic layered
media (periodic stacks), which can be viewed as one dimensional photonic
crystals. According to \cite{PRB03}, at normal incidence, the frozen mode
regime in a periodic stack can only occur if some of the layers display
sufficiently strong circular birefringence (Faraday rotation). In addition,
each unit cell of the periodic layered array must contain at least two layers
with significant and misaligned in-plane anisotropy. If the above conditions
are not met, the electromagnetic dispersion relation of the periodic stack
cannot develop stationary inflection point (\ref{SIP}) and, therefore, cannot
support the frozen mode regime at normal incidence. At the microwave frequency
range, one can find a number of materials meeting the above requirements. But
at infrared and optical frequencies, the circular birefringence of known
transparent magnetic materials becomes too small to support a robust frozen
mode regime \cite{PRB03}. Since our prime interest here is with optics, we
will explore the "non-magnetic" approach proposed in \cite{PRE03}.

According to \cite{PRE03}, the frozen mode regime can occur in nonmagnetic
periodic stacks with special configurations requiring some layers to display
appreciable oblique (neither in-plane, nor axial) anisotropy. On the other
hand, at optical frequencies, all commercially available anisotropic
dielectrics display substantially weaker anisotropy, compared to what would be
the optimal value. According to \cite{PRE03}, too weak anisotropy can push the
frozen mode frequency too close to the nearest band edge, resulting in almost
total reflectance of the incident light. The high reflectance of the slab, in
turn, implies very low efficiency of conversion of the incident light into the
slow mode. In this paper we show that in fact, the negative effect of the weak
anisotropy on the frozen mode regime can be completely overcome by proper
design of the layered structure. As the result, a robust frozen mode regime at
optical frequencies can be achieved in periodic stacks incorporating real
anisotropic materials such as yttrium vanadate, lithium niobate, and the like,
where the dielectric anisotropy is one-two orders of magnitude short of the
"optimal" value. The idea is to choose the parameters of the periodic stack so
that a stationary inflection point associated with the frozen mode regime
develops at higher photonic bands. For a given frequency range, this requires
thicker dielectric layers, which could be an additional practical advantage. A
side effect of using higher photonic bands is that the effective bandwidth of
the frozen mode regime appears to be narrower, compared to the case of
hypothetical materials with much stronger anisotropy used in \cite{PRE03} for
numerical simulations.

The practical development of frozen mode devices from such commodity materials
could lead to revolutionary advances in optical computing, sensing, and
information processing. When practically realized, such frozen mode structures
would enable significant advances in all-optical information storage and
processing (such as optical memory and buffer elements, optical delay lines)
as well as optical sensing, lasing, and nonlinear optics.

The paper is organized as follows. In the next section, we discuss in general
terms the phenomenon of axially frozen mode. The detailed analysis of the
mathematical aspects of the phenomenon can be found in \cite{PRE03}. Then, in
section III, using a specific example of a periodic array incorporating
yttrium vanadate, we demonstrate how a robust frozen mode regime at optical
frequencies can be achieved in a practical setting involving weakly
anisotropic materials. Finally, in Appendix, we briefly overview the
electrodynamics of lossless layered media, introducing basic notations,
definitions, and assumptions used in our computations.

\section{Axially frozen mode regime at oblique incidence}

According to \cite{PRB03,PRE03}, in nonmagnetic periodic stacks, the simplest
version of the frozen mode regime described in the introduction is impossible.
Still, a more general phenomenon referred to as the \emph{axially frozen mode
regime} can occur. This section starts with a brief general discussion of the
phenomenon. Then we turn to the particular case of a periodic stack
incorporating yttrium vanadate layers. The reason we have chosen this
particular material is because its optical properties is very similar to those
of other common anisotropic dielectrics transparent at optical wavelength.

\subsection{Basic definitions}

\begin{figure}[tbph]
\includegraphics[scale=0.9, viewport= -100 0 300 220, clip]{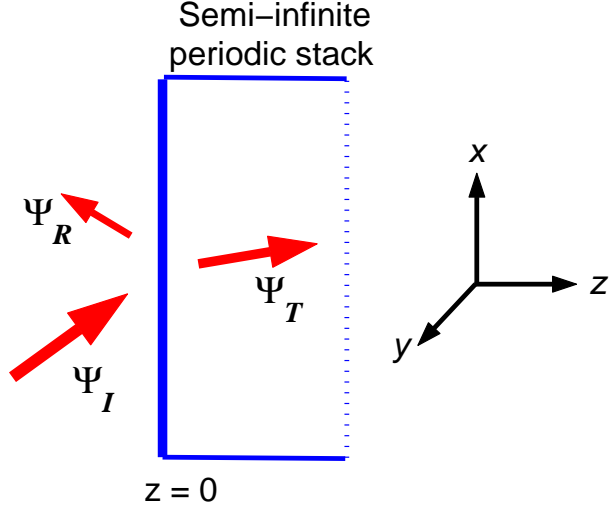}
\caption{Light incident on a semi-infinite periodic layered medium. The arrows
show the energy fluxes of the incident, reflected and transmitted waves,
respectively. The transmitted wave $\Psi_{T}$ is a superposition of two Bloch
eigenmodes, each of which can be either propagating or evanescent. Only
propagating modes can transfer electromagnetic energy in the $z$ direction.}
\label{SIS}
\end{figure}

Consider a monochromatic plane wave obliquely incident on a periodic
semi-infinite stack, as shown in Fig. \ref{SIS}.%

\begin{figure}[tbph]
\includegraphics[scale=0.9, viewport= -50 0 400 420, clip]{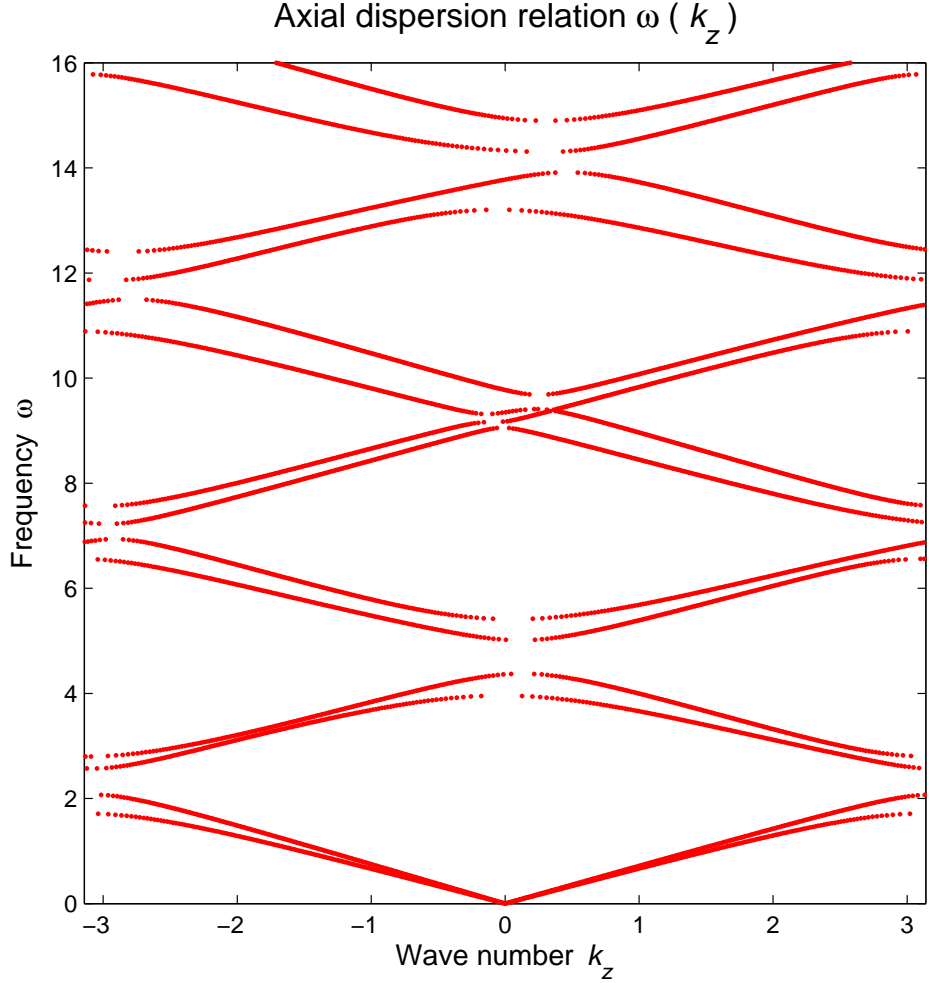}
\caption{The axial dispersion relation $\omega(k_{z})$ of the periodic stack
in Fig. 4 at fixed direction $\vec{n}$ of light incidence ($n_{x}%
=n_{y}=-0.493489$). Small spectral asymmetry and small branch separation are
due to weak anisotropy of yttrium vanadate. Both the asymmetry and the branch
separation are more prononced in upper spectral bands. The wave number $k_{z}$
and the frequency $\omega$ are given in units of $1/L$ and $c/L$, respectively.}
\label{DRn_YV}
\end{figure}

\begin{figure}[tbph]
\includegraphics[scale=0.9, viewport= 0 0 500 420, clip]{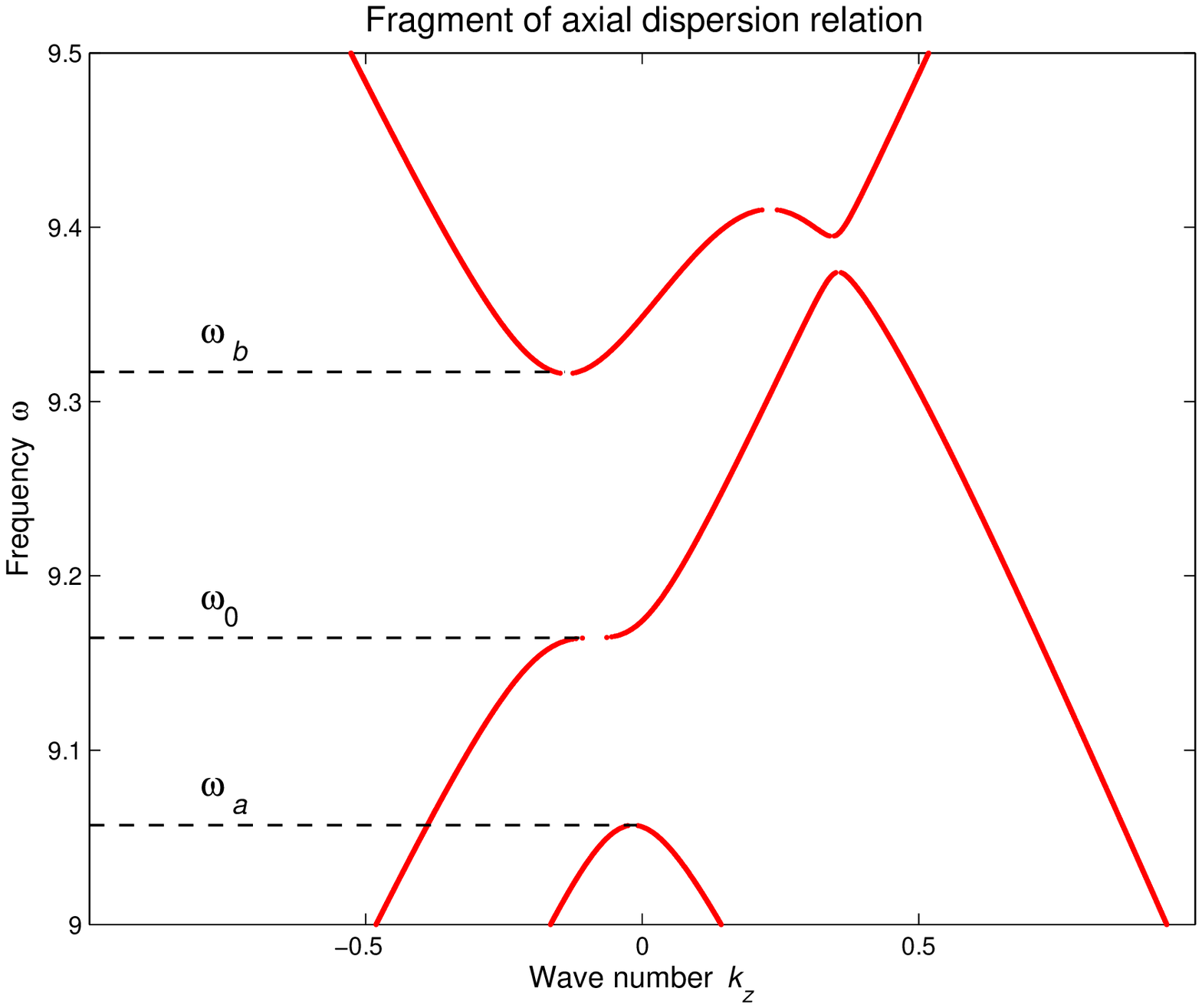}
\caption{A fragment of the axial dispersion relation $\omega\left(k_{z}\right)  $ in Fig. 5, which includes the axial stationary inflection
point at $\omega_{0}=9.164450223$.}
\label{DRn_YV_f}
\end{figure}

Let $\Psi_{I}$, $\Psi_{R}$ and $\Psi_{T}$\ denote the incident, reflected and
transmitted waves, respectively. Due to the boundary conditions (\ref{BC}),
all three waves $\Psi_{I}$, $\Psi_{R}$ and $\Psi_{T}$ \ must be assigned the
same pair of tangential components $k_{x},k_{y}$ of the respective wave
vector\cite{LLEM 84,Yariv}%
\begin{equation}
\left(  \vec{k}_{I}\right)  _{x}=\left(  \vec{k}_{R}\right)  _{x}=\left(
\vec{k}_{T}\right)  _{x},\ \ \left(  \vec{k}_{I}\right)  _{y}=\left(  \vec
{k}_{R}\right)  _{y}=\left(  \vec{k}_{T}\right)  _{y}, \label{k_x k_y}%
\end{equation}
while their axial (normal) components $k_{z}$ can be different. Hereinafter,
the symbol $k_{z}$ will refer only to the transmitted Bloch waves propagating
inside the semi-infinite slab%
\begin{equation}
\text{Inside periodic stack (at }z>0\text{): \ \ }\vec{k}=\left(  k_{x}%
,k_{y},k_{z}\right)  . \label{k=kx,ky,kz}%
\end{equation}
The value $k_{z}$ is defined up to a multiple of $2\pi/L$ (the Brillouin
zone), where $L$ is the period of the layered structure. For given
$k_{x},k_{y}$ and $\omega$, the value $k_{z}$ is found by solving the
time-harmonic Maxwell equations (\ref{MEz}) in periodic medium, that will be
done in the following sections. The result can be represented as the
\emph{axial dispersion relation}, which gives the correspondence between
$\omega$ and $k_{z}$ at fixed $k_{x},k_{y}$%
\begin{equation}
\omega=\omega\left(  k_{z}\right)  ,\text{\ \ at fixed }k_{x},k_{y}\text{.}
\label{ADR,k}%
\end{equation}
It can be more convenient to define the axial dispersion relation as the
correspondence between $\omega$ and $k_{z}$ at fixed direction $\vec{n}$ of
incident light propagation%
\begin{equation}
\omega=\omega\left(  k_{z}\right)  ,\text{\ \ at fixed }n_{x},n_{y},
\label{ADR,n}%
\end{equation}
where the unit vector $\vec{n}$ can be expressed in terms of the tangential
components (\ref{k_x k_y}) of the wave vector%
\begin{equation}
n_{x}=k_{x}c/\omega,\ \ n_{y}=k_{y}c/\omega,\ \ n_{z}=\sqrt{1-\left(
n_{x}^{2}+n_{y}^{2}\right)  }. \label{n(k)}%
\end{equation}
Examples of axial dispersion relation (\ref{ADR,n}) are presented in Figs.
\ref{DRn_YV} and \ref{DRn_YV_f}.

\begin{figure}[tbph]
\includegraphics[scale=0.9, viewport= 0 0 400 220, clip]{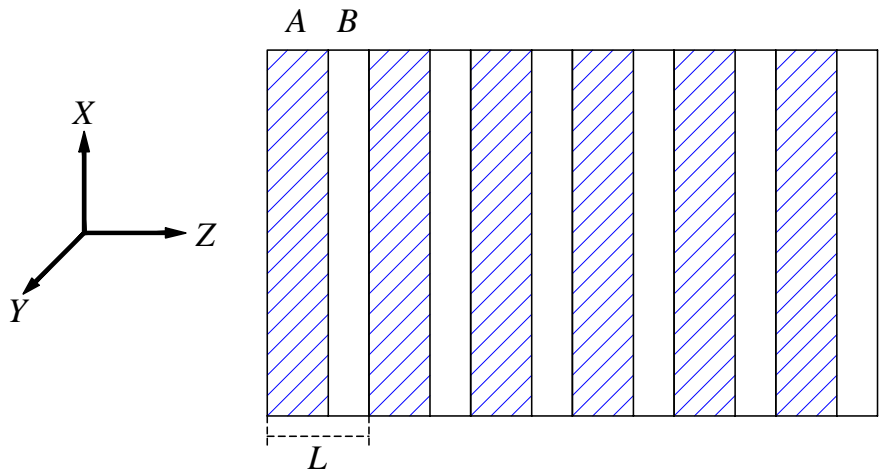}
\caption{Periodic array of anisotropic dielectric layers ($A$) separated by
gaps ($B$). The anisotropy axis of the dielectric material (the tetragonal
axis, in the case of yttrium vanadate) makes an oblique angle with the $z$
direction, normal to the layers. The stack parameters used in our numerical
simulations are specified in formulae (30) and (31).}
\label{StackAB}
\end{figure}

The transmitted electromagnetic field $\Psi_{T}$ inside the periodic layered
medium is not a single Bloch mode, but it is a superposition of two Bloch
modes with different polarizations and different values of $k_{z}$. Of course,
the tangential components $k_{x},k_{y}$ are the same for either transmitted
Bloch mode and the incident wave, as stated by Eq. (\ref{k_x k_y}). Generally,
there are three possibilities (see the details in \cite{PRE03}):

\begin{enumerate}
\item Both Bloch components of the transmitted wave $\Psi_{T}$ are propagating
modes, which means that the respective values of $k_{z}$ are real. For
example, at $\omega<\omega_{a}$ and $\omega>\omega_{b}$ in Fig. \ref{DRn_YV_f}
we have two Bloch modes propagating inside the slab with two different group
velocities $u_{z}>0$ (double refraction). Note that propagating modes with
$u_{z}<0$, as well as evanescent modes with $\Im \left(
k_{z}\right)  <0$, do not contribute to the transmitted wave $\Psi_{T}$ inside
the semi-infinite stack in Fig. \ref{SIS}.

\item Both Bloch components of $\Psi_{T}$ are evanescent, which implies that
the respective values of $k_{z}$ are complex with $\Im \left(
k_{z}\right)  >0$. In particular, this is the case if the frequency $\omega$
falls into a photonic band gap (for example, at $\omega>\omega_{g}$ in Fig.
\ref{DR_PRE03}). In such a case, the incident wave is totally reflected back
to space.

\item Of particular interest is the case when one of the Bloch components of
the transmitted wave $\Psi_{T}$ is a propagating mode (with $u_{z}>0$), while
the other is an evanescent mode (with $\Im \left(  k_{z}\right)
>0$)%
\begin{equation}
\Psi_{T}=\Psi_{pr}+\Psi_{ev}. \label{Psi_T=ex+ev}%
\end{equation}
This is the case at the frequency range%
\begin{equation}
\omega_{a}<\omega<\omega_{b} \label{wa<w0<wb}%
\end{equation}
in Fig. \ref{DRn_YV_f}. As the distance $z$ from the slab/vacuum interface
increases, the evanescent contribution $\Psi_{ev}$ decays as $\exp\left(
-z\Im \left(  k_{z}\right)  \right)  $, and the resulting
transmitted wave $\Psi_{T}$ turns into a single propagating Bloch eigenmode
$\Psi_{pr}$.
\end{enumerate}

\begin{figure}[tbph]
\includegraphics[scale=0.9, viewport= 0 0 500 200, clip]{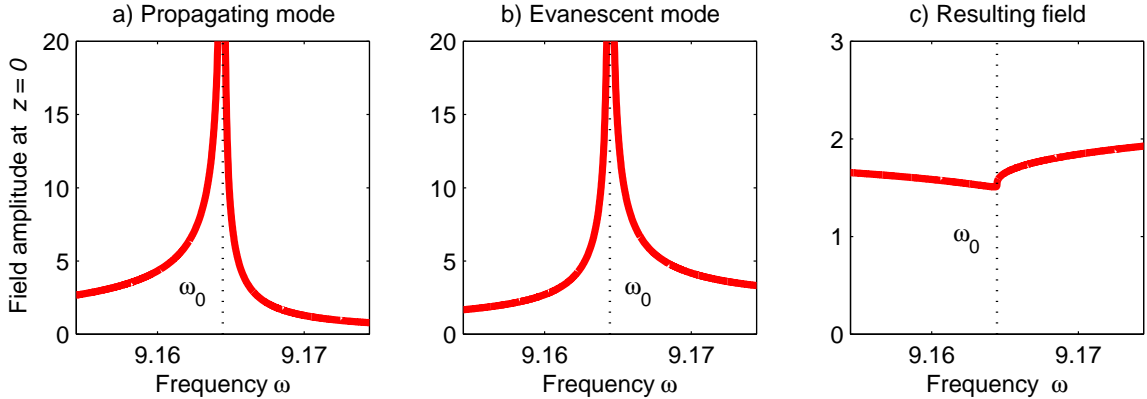}
\caption{Destructive interference of the propagating and evanescent
contributions to the resulting field $\Psi_{T}$ at the slab/vacuum interface
under the frozen mode regime: a) amplitude $\left\vert \Psi_{pr}\left(
0\right)  \right\vert ^{2}$ of the propagating component, b) amplitude
$\left\vert \Psi_{ev}\left(  0\right)  \right\vert ^{2}$ of the evanescent
component, c) resulting field amplitude $\left\vert \Psi_{T}\left(  0\right)
\right\vert ^{2}$. The incident light has unity amplitude and TE polarization.}
\label{AM0_YV_TE}
\end{figure}

Similarly to the case (\ref{SIP}) of regular frozen mode, the \emph{axially
frozen mode} is associated with the \emph{axial stationary inflection point}
defined as
\begin{equation}
\text{\ at \ }\omega=\omega_{0}\text{:\ \ }\frac{\partial\omega}{\partial
k_{z}}=0;\;\frac{\partial^{2}\omega}{\partial k_{z}^{2}}=0;\;\frac
{\partial^{3}\omega}{\partial k_{z}^{3}}\neq0.\label{ASIP}%
\end{equation}
The regular stationary inflection point (\ref{SIP}) is a particular case of
(\ref{ASIP})$.$ Example of axial dispersion relation displaying such a
singularity is shown in Fig. \ref{DRn_YV_f}. In the vicinity of $\omega_{0}$
in Eq. (\ref{ASIP}), the electromagnetic field $\Psi_{T}$ inside the slab is a
superposition (\ref{Psi_T=ex+ev}) of one propagating and one evanescent Bloch
components. As the frequency $\omega$ approaches the critical point
(\ref{ASIP}), both contributions grow sharply, while remaining nearly equal
and opposite in sign near the slab boundary (at $z=0$), as illustrated in Fig.
\ref{AM0_YV_TE}.%

\begin{figure}[tbph]
\includegraphics[scale=0.9, viewport= 0 0 500 200, clip]{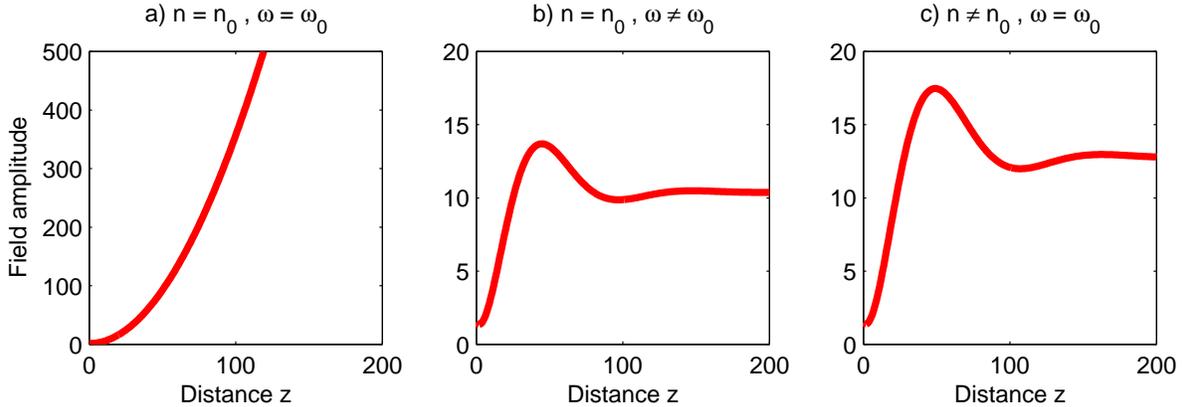}
\caption{The Amplitude $\left\vert \Psi\left(  z\right)  \right\vert ^{2}$ of
the resulting field as function of the distance $z$ from the slab boundary: a)
the exact point of the frozen mode regime, b) the frequency is deviated from
$\omega_{0}$ by $10^{-3}c/L$, c) the direction of light incidence is deviated
from $n_{0}$ in (\ref{n0}) by $10^{-4}$. The distance $z$ is given in units of $L$. The
incident light has unity amplitude and TE polarization.}
\label{AMz_YV_TE}
\end{figure}

Due to the destructive interference at the slab boundary, the resulting
electromagnetic field at $z=0$ is small enough to satisfy the boundary
condition (\ref{BC}). As the distance $z$ from the slab boundary increases,
the evanescent component $\Psi_{ev}$ decays exponentially, while the amplitude
of the propagating component $\Psi_{pr}$ remains constant and large, as shown
in Figs. \ref{AMz_YV_TE}\emph{b} and \ref{AMz_YV_TE}\emph{c}.%

\subsection{Energy density and energy flux of the axially frozen mode}

Let $\vec{S}_{I}$, $\vec{S}_{R}$ and $\vec{S}_{T}$ be the energy fluxes of the
incident, reflected and transmitted waves, respectively. Within the frequency
range (\ref{wa<w0<wb}), which includes the critical point (\ref{ASIP}), the
transmitted wave $\Psi_{T}$ is a superposition (\ref{Psi_T=ex+ev}) of
propagating and evanescent components. Only the propagating component
$\Psi_{pr}$ is responsible for the axial energy flux $\left(  \vec{S}%
_{T}\right)  _{z}$.

The axial energy flux can also be expressed in terms of the axial component
$u_{z}$ of the propagating mode group velocity and the energy density $W_{0}$
associated with $\Psi_{pr}$%
\begin{equation}
\left(  \vec{S}_{T}\right)  _{z}=W_{0}u_{z}\propto\left\vert \Psi
_{pr}\right\vert ^{2}u_{z},\label{Sz=Wuz}%
\end{equation}
The quantity $W_{0}$ in Eq. (\ref{Sz=Wuz}) can be interpreted as the
electromagnetic energy density far from the slab interface, where the
electromagnetic field $\Psi_{T}$ reduces to its propagating component
$\Psi_{pr}$. In the vicinity of the axial stationary inflection point
(\ref{ASIP}), the energy density $W_{0}$, associated with the axially frozen
mode, diverges, while $u_{z}\rightarrow0$. As a result, the vanishingly small
$u_{z}$ in Eq. (\ref{Sz=Wuz}) is offset by a very large value of $W_{0}$. The
theoretical analysis of the next section shows that the axial energy flux
$\left(  \vec{S}_{T}\right)  _{z}$ in Eq. (\ref{Sz=Wuz}), along with the slab
transmittance $\tau$ in (\ref{tau, rho}), remain finite even at $\omega
=\omega_{0}$, where the axial component of the group velocity vanishes%
\begin{equation}
\left(  \vec{S}_{T}\right)  _{z}>0\text{,}\;\text{if}\;u_{z}=0.\label{Sz > 0}%
\end{equation}

\begin{figure}[tbph]
\includegraphics[scale=0.9, viewport= 0 0 500 450, clip]{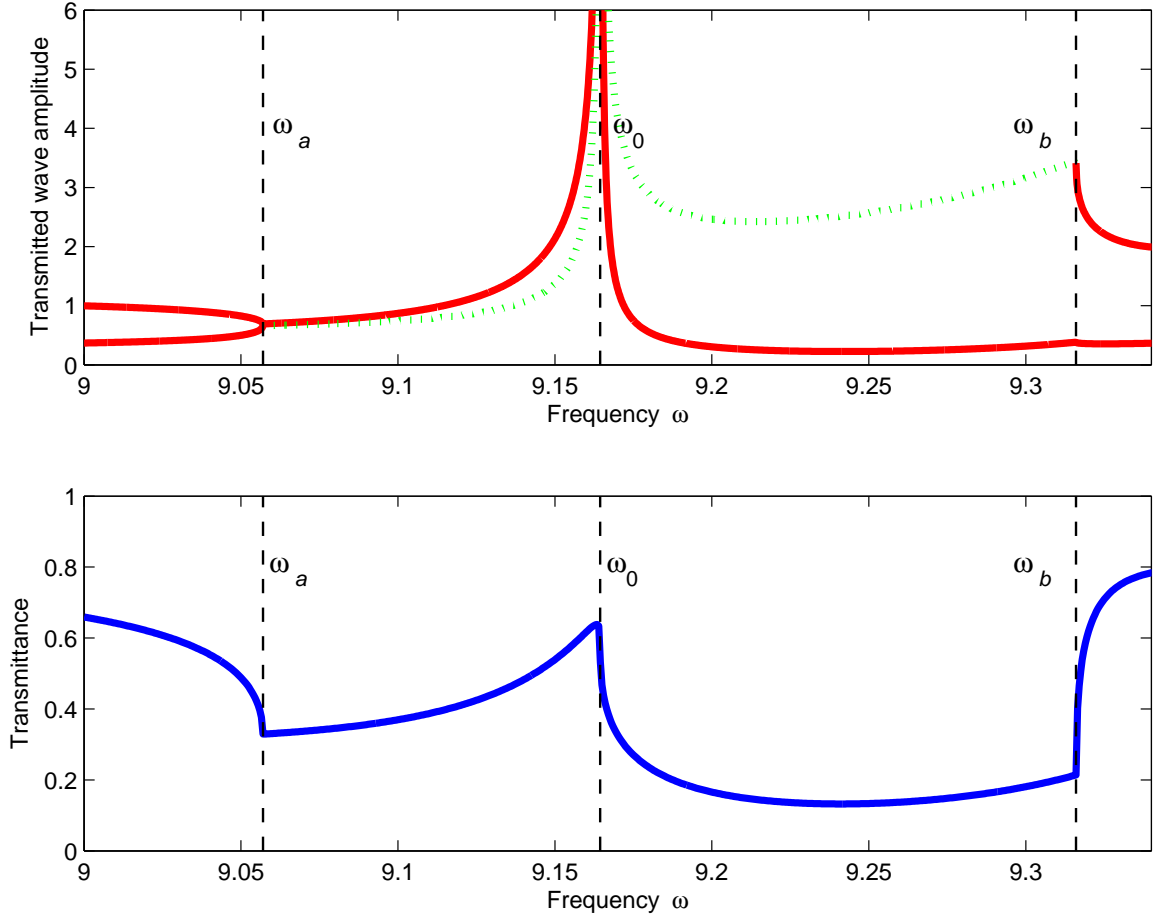}
\caption{At the top: the red solid line shows the amplitude $\left\vert
\Psi_{pr}\right\vert ^{2}$ of the propagating Bloch component of the
transmitted wave $\Psi_{T}$ from Eq. (11). The dotted green line shows the
amplitude $\left\vert \Psi_{ev}\right\vert ^{2}$ of the evanescent component.
Both contributions to $\Psi_{T}$ diverge at the frozen mode frequency
$\omega_{0}$. At $z\gg L$, the evanescent component decays and $\left\vert
\Psi_{pr}\right\vert ^{2}$ (the solid line) represents the resulting amplitude
$\left\vert \Psi_{T}\right\vert ^{2}$ of the transmitted wave. By contrast, at
$\omega<\omega_{a}$ and $\omega>\omega_{b}$, the transmitted wave $\Psi_{T}$
is a superposition of two propagating components (double refraction). At the
bottom: the slab transmittance $\tau$ vs. frequency $\omega$. At
$\omega=\omega_{0}$, the transmittance remains finite. The characteristic
frequencies $\omega_{a}$, $\omega_{0}$ and $\omega_{b}$ are explained in Fig.
6. The incident wave has TE polarization and unity amplitude.}
\label{tr_AM_TE}
\end{figure}

\begin{figure}[tbph]
\includegraphics[scale=0.9, viewport= 0 0 500 450, clip]{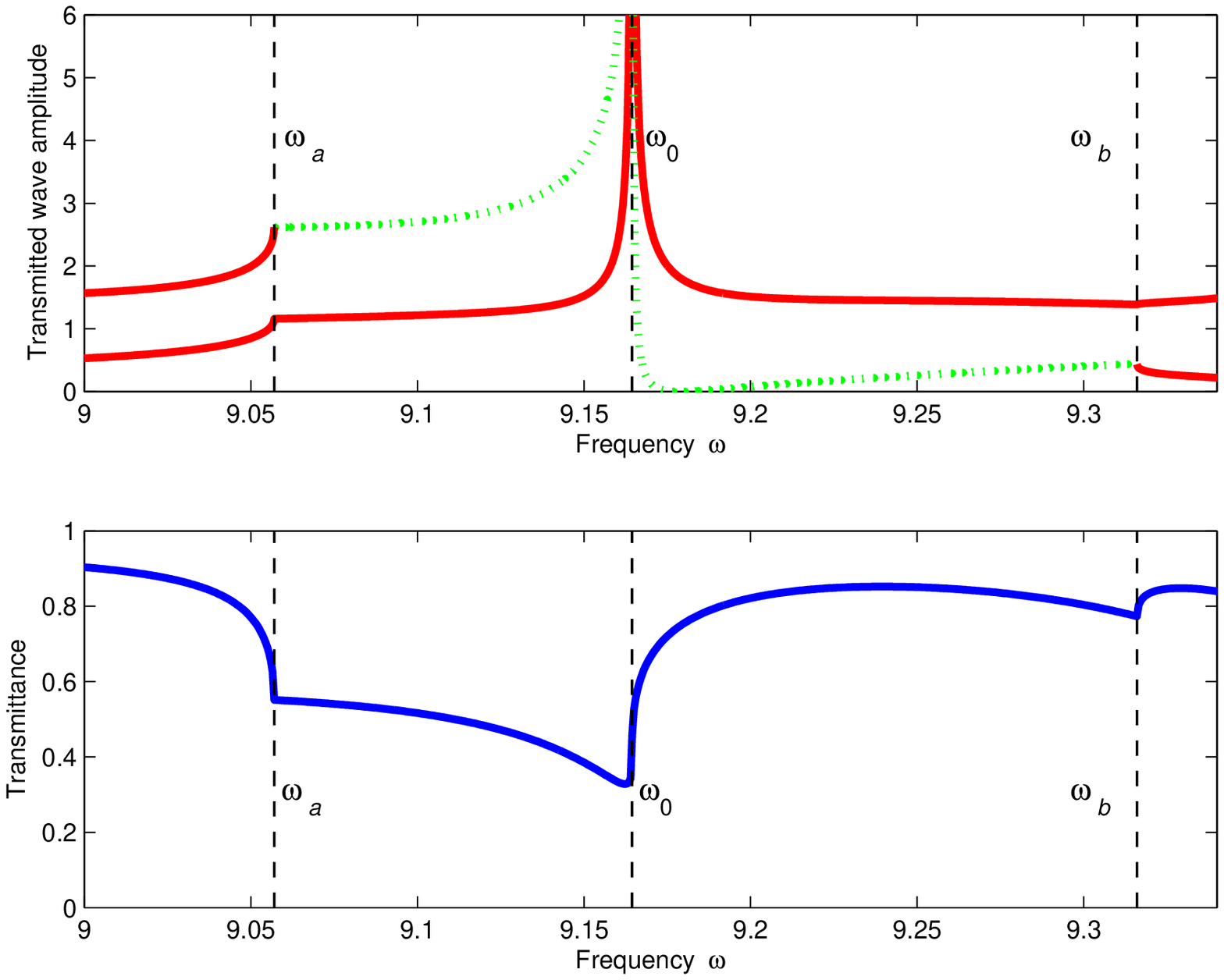}
\caption{The same as in Fig. 9, but the incident wave polarization is now TH.}
\label{tr_AM_TH}
\end{figure}

The energy conservation consideration allows to find the asymptotic frequency
dependence of the amplitude $\left\vert \Psi_{pr}\right\vert ^{2}$ of the
axially frozen mode in the vicinity of the critical point (\ref{ASIP}).
Indeed, in the vicinity of $\omega_{0}$, the axial dispersion relation can be
approximated by a cubic parabola%
\begin{equation}
\omega-\omega_{0}\approx\frac{1}{6}\omega_{0}^{\prime\prime\prime}\left(
k_{z}-k_{0}\right)  ^{3},\ \ \ \omega_{0}^{\prime\prime\prime}=\left(
\frac{\partial^{3}\omega}{\partial k_{z}^{3}}\right)  _{\vec{k}=\vec{k}_{0}%
}.\label{w=k^3}%
\end{equation}
The $z$ component of the group velocity is%
\begin{equation}
u_{z}=\frac{\partial\omega}{\partial k_{z}}\approx\frac{1}{2}\omega
_{0}^{\prime\prime\prime}\left(  k_{z}-k_{0}\right)  ^{2}\approx\frac{6^{2/3}%
}{2}\left(  \omega_{0}^{\prime\prime\prime}\right)  ^{1/3}\left(
\omega-\omega_{0}\right)  ^{2/3}.\label{u=w^2/3}%
\end{equation}
The Eq. (\ref{u=w^2/3}) together with (\ref{Sz=Wuz}) yield the following
asymptotic expression for the energy density $W_{0}$ associated with the
frozen mode%
\begin{equation}
W_{0}\approx\frac{2}{6^{2/3}}\left(  \vec{S}_{T}\right)  _{z}\left(
\omega_{0}^{\prime\prime\prime}\right)  ^{-1/3}\left(  \omega-\omega
_{0}\right)  ^{-2/3},\label{W(ST)}%
\end{equation}
or, equivalently,%
\begin{equation}
W_{0}\approx\frac{2}{6^{2/3}}\tau\left(  \vec{S}_{I}\right)  _{z}\left(
\omega_{0}^{\prime\prime\prime}\right)  ^{-1/3}\left(  \omega-\omega
_{0}\right)  ^{-2/3}.\label{W(w)}%
\end{equation}
where $\vec{S}_{I}$ is the fixed energy flux of the incident wave and $\tau$
is the transmittance coefficient defined in (\ref{tau, rho}). Remarkably, the
transmittance $\tau$ of the semi-infinite slab remains finite in the vicinity
of the frozen mode frequency $\omega_{0}$, as seen in Figs. \ref{tr_AM_TE} and
\ref{tr_AM_TH}. This implies that the electromagnetic energy density $W_{0}$
associated with the frozen mode, as well as its amplitude $\left\vert
\Psi_{pr}\right\vert ^{2}$, diverge as $\omega\rightarrow\omega_{0}$. In Figs.
\ref{tr_AM_TE} and \ref{tr_AM_TH} such a behavior is illustrated for two
different incident light polarizations.%

In reality, under the axial frozen mode regime, the field amplitude inside the
slab is limited by various physical factors mentioned earlier in this paper.
Still, with all these limitations in place, the normal energy flux $\left(
\vec{S}_{T}\right)  _{z}$ remains finite and comparable with that of the
incident wave. The latter implies that a substantial portion of the incident
wave is converted into the axially frozen mode with drastically enhanced
amplitude and nearly zero axial component $u_{z}$ of the group velocity. In
many respects, the phenomenon of axial frozen mode is similar to its
particular case, the regular frozen mode, associated with the regular
stationary inflection point (\ref{SIP}).

\subsection{Physical conditions for the axially frozen mode regime in layered
media}

The physical conditions under which a non-magnetic layered structure can
support the (axial) frozen mode regime can be grouped in two categories. The
first one comprises several symmetry restrictions. The second category
includes some basic qualitative recommendations which would ensure the
robustness of the frozen mode regime, provided that the symmetry conditions
for the regime are met. In what follows we briefly reiterate those conditions
and then show how they apply to periodic stacks incorporating some real
dielectric materials.

\subsubsection{Symmetry conditions}

There are two fundamental necessary conditions for the frozen mode regime. The
first one is that the Bloch dispersion relation $\omega\left(  \vec{k}\right)
$ in the periodic layered medium must display the so-called \emph{axial
spectral asymmetry}%
\begin{equation}
\omega\left(  k_{x},k_{y},k_{z}\right)  \neq\omega\left(  k_{x},k_{y}%
,-k_{z}\right)  . \label{asymm}%
\end{equation}
As shown in \cite{PRE03}, this condition is necessary for the existence of the
axial stationary \ inflection point (\ref{ASIP}) in the electromagnetic
dispersion relation of an arbitrary periodic layered medium.

The second necessary condition is that for the given direction $\vec{k}$ of
wave propagation, the Bloch eigenmodes $\Psi_{\vec{k}}$ with different
polarizations must have the same symmetry. In the case of oblique propagation
in periodic layered media, the latter condition implies that for the given
$\vec{k}$, the Bloch eigenmodes\ are neither TE nor TM%
\begin{equation}
\Psi_{\vec{k}}\text{ \ is neither TE nor TM.} \label{TE/TM}%
\end{equation}

The condition (\ref{asymm}) imposes certain restrictions on (i) the point
symmetry group $G$ of the periodic layered array and (ii) on the direction
$\vec{k}$ of the transmitted wave propagation inside the layered medium. While
the condition (\ref{TE/TM}) may impose some additional restriction on the
direction of $\vec{k}$.

The restriction on the symmetry of the periodic stack following from the
requirement (\ref{asymm}) of the axial spectral asymmetry is%
\begin{equation}
m_{z}\notin G\text{ \ and \ }2_{z}\notin G. \label{asymC2}%
\end{equation}
where $m_{z}$ is the mirror plane parallel to the layers, $2_{z}$ is the
2-fold rotation about the $z$ axis. An immediate consequence of the criterion
(\ref{asymC2}) is that least one of the alternating layers of the periodic
stack must be an anisotropic dielectric material with nonzero $\varepsilon
_{xz}$ and/or $\varepsilon_{yz}$, where the $z$ direction is normal to the
layers%
\begin{equation}
\varepsilon_{xz}\neq0\text{ \ \ and/or \ \ }\varepsilon_{yz}\neq0
\label{eps<>0}%
\end{equation}
Otherwise, the operation $2_{z}$ will be present in the symmetry group $G$ of
the periodic stack.

The condition (\ref{asymm}) also imposes a restriction on the direction of
wave propagation. Specifically, the Bloch wave vector $\vec{k}$ must be
oblique to the stack layers, which means that $\vec{k}$ is neither parallel,
nor perpendicular to the $z$ direction%
\[
k_{x}\neq0\text{ \ \ and/or \ \ }k_{y}\neq0.
\]
The latter condition implies that the frozen mode regime cannot occur at
normal incidence, regardless of the periodic stack geometry and composition.
While the condition (\ref{eps<>0}) implies that at least one of the
alternating layers must be cut at an oblique angle relative to the principle
axes of its permittivity tensor. If either of the above two conditions is not
satisfied, the dispersion relation $\omega\left(  \vec{k}\right)  \ $will be
axially symmetric%
\begin{equation}
\omega\left(  k_{x},k_{y},k_{z}\right)  =\omega\left(  k_{x},k_{y}%
,-k_{z}\right)  , \label{18}%
\end{equation}
which rules out the possibility of stationary inflection point and the frozen
mode regime.

If all the above necessary conditions are met, then the (axial) frozen mode
regime is, at least, not forbidden by symmetry. More details on the symmetry
aspects of the frozen mode regime can be found in \cite{PRE03}, Section II and
\cite{PRB03}, Sections I and II.

\subsubsection{Additional physical requirements}

In practice, as soon as the symmetry conditions are met, one can almost
certainly achieve the (axial) frozen mode regime at any desirable frequency
$\omega$ within certain frequency range. The frequency range
is determined by the layers thicknesses and the dielectric
materials used, while a specific value of $\omega$ within the range can be
selected by the direction $\vec{n}$ of the light incidence. The problem is
that unless the physical parameters of the stack layers lie within a certain
range, the effects associated with the frozen mode regime can be insignificant
or even practically undetectable. The basic guiding principle in choosing
appropriate layer materials is "moderation". As soon as the "moderation"
principle is observed, one can almost certainly achieve the frozen mode regime
at prescribed frequency by choosing the right direction of light incidence.
Specifically, those "moderation" conditions include:

\begin{enumerate}
\item It is desirable that the ratio%
\begin{equation}
\frac{\Delta n}{n} \label{e11/ep33}%
\end{equation}
of the birefringence and the refractive index of the material of the anisotropic
layers lies somewhere between 2 and 10.
If the anisotropy is extremely strong or too weak, the Bloch waves with
different polarizations become virtually separated, which excludes the
possibility of a robust frozen mode regime. For example, in the practically
important case of extremely small anisotropy, the two transmitted Bloch waves
can be approximately classified as TE and TM modes, which is incompatible with
the symmetry condition (\ref{TE/TM}) for the frozen mode regime.

\item The dielectric contrast of the adjacent layers $A$ and $B$ should be
significant, but not extreme. The ratio $n_{A}/n_{B}$ anywhere between 1.5 and
20 would be appropriate. In addition, the dielectric contrast between the
layers should match the ratio (\ref{e11/ep33}) in the anisotropic layers:
weaker anisotropy would require weaker dielectric contrast between the layers.

\item Typical layer thickness should be of the order of $\lambda/4n$, where
$\lambda=2\pi/\omega$ is the light wavelength in vacuum and $n$ is the
respective refractive index. In reality, the acceptable layer thickness can
differ from $\lambda/4n$ by several times either way. But too thick layers
would push the stationary inflection point to high-order frequency bands,
while too thin layers would exclude the possibility of the frozen mode regime
at the prescribed frequency range.
\end{enumerate}

The biggest challenge at optical frequencies poses the first condition,
because most of the commercially available optical anisotropic crystals have
the ratio $\Delta n/n$ of only about 0.1. According to \cite{PRE03}, this
would push the axial stationary inflection point (\ref{ASIP}) very close to
the photonic band edge and make the photonic crystal almost 100\% reflective.
This indeed would be the case if we tried to realize the frozen mode regime at
the lowest frequency band. But, in the next section we show that one can
successfully solve this problem by moving to a higher frequency band. So, a
robust axially frozen mode regime with almost complete conversion of the
incident light into the frozen mode can be achieved with the commercially
available anisotropic dielectric materials displaying the birefringence ratio
(\ref{e11/ep33}) by one-two orders of magnitude smaller, compared to the
optimal value used in \cite{PRE03} for numerical simulations. The drawback
though is that using higher photonic frequency bands narrows the bandwidth of
the frozen mode regime by roughly an order of magnitude.

\section{Periodic stack incorporating Yttrium Vanadate layers}

The simplest non-magnetic periodic stack satisfying the symmetry conditions
(\ref{asymC2}) and (\ref{eps<>0}) and, therefore, capable of supporting the
axial frozen mode regime, is shown in Fig. \ref{StackAB}. It is composed of
anisotropic $A$ layers separated by empty gaps. The anisotropic dielectric
layers must display nonzero components $\varepsilon_{xz}$ and/or
$\varepsilon_{yz}$. The simplest choice for the respective permittivity tensor
is%
\begin{equation}
\hat{\varepsilon}_{A}=\left[
\begin{array}
[c]{ccc}%
\varepsilon_{xx} & 0 & \varepsilon_{xz}\\
0 & \varepsilon_{yy} & 0\\
\varepsilon_{xz} & 0 & \varepsilon_{zz}%
\end{array}
\right]  \label{eps A}%
\end{equation}
where the $z$ axis coincides with the normal to the layers.

Yttrium vanadate is a tetragonal dielectric with the permittivity tensor,
$\hat{\varepsilon}_{YV}$, at $\lambda=1550$ nm \cite{eps}%
\begin{equation}
\hat{\varepsilon}_{YV}=\left[
\begin{array}
[c]{ccc}%
\varepsilon_{11} & 0 & 0\\
0 & \varepsilon_{22} & 0\\
0 & 0 & \varepsilon_{33}%
\end{array}
\right]  =\left[
\begin{array}
[c]{ccc}%
4.62 & 0 & 0\\
0 & 4.62 & 0\\
0 & 0 & 3.78
\end{array}
\right]  \label{epsYV}%
\end{equation}
where the Cartesian axis $x_{3}$ is chosen parallel to the crystallographic
axis $C_{4}$. In order to achieve a non-zero $\varepsilon_{xz}$ component one
has to rotate the tensor (\ref{epsYV}) about the $y$ axis by an angle $\theta
$, different from $0$ and $\pi/2$ \cite{Ballato}. The result of the rotation
is%
\begin{equation}
\hat{\varepsilon}_{A}=\left[
\begin{array}
[c]{ccc}%
\varepsilon_{11}\cos^{2}\theta+\varepsilon_{33}\sin^{2}\theta & 0 & \left(
\varepsilon_{11}-\varepsilon_{33}\right)  \cos\theta\sin\theta\\
0 & \varepsilon_{22} & 0\\
\left(  \varepsilon_{11}-\varepsilon_{33}\right)  \cos\theta\sin\theta & 0 &
\varepsilon_{33}\cos^{2}\theta+\varepsilon_{11}\sin^{2}\theta
\end{array}
\right]  . \label{eps A rot}%
\end{equation}
For numerical simulations we can choose, for instance,%
\begin{equation}
\theta=\pi/4. \label{Pi/4}%
\end{equation}
In this case, Eq. (\ref{eps A rot}) together with (\ref{epsYV}) yield%
\begin{equation}
\hat{\varepsilon}_{A}=\left[
\begin{array}
[c]{ccc}%
4.20 & 0 & 4.42\\
0 & 4.62 & 0\\
4.42 & 0 & 4.20
\end{array}
\right]  \label{epsA}%
\end{equation}
which is compatible with the required form (\ref{eps A}).

Let $d_{A}$ and $d_{B}$ denote the thickness of the $A$ layers and the
thickness of the gaps between them, respectively. For our numerical
simulations we can choose%
\begin{equation}
d_{A}=d_{B}=L/2, \label{dA=dB}%
\end{equation}
where $L$ is the period of the layered array in Fig. \ref{StackAB}.

Let us reiterate that the parameters (\ref{Pi/4}) and (\ref{dA=dB}) are chosen
at random. In practice, we can always adjust them so that the stack suits
specific practical requirements. The structural period $L$ should be chosen so
that the frozen mode regime occurs within a prescribed frequency range. Then,
the direction of light incidence can be adjusted so that the axial frozen mode
regime occurs exactly at a prescribed frequency $\omega_{0}$.

Symmetry arguments similar to those presented in \cite{PRE03} show that in the
case of the periodic array in Fig. \ref{StackAB}, the necessary conditions
(\ref{asymC2}) and (\ref{TE/TM}) for the frozen mode regime are satisfied only
if the direction $\vec{n}$ of the light incidence lies neither in the $x-z$,
\ nor $y-z$ plane%
\begin{equation}
n_{x}\neq0\text{ \ and \ }n_{y}\neq0. \label{nx,ny <> 0}%
\end{equation}

\subsection{Electromagnetic properties in the vicinity of axial frozen mode
regime}

In Fig. \ref{DRn_YV} we presented the typical axial dispersion relation
$\omega(k_{z})$ of the periodic stack in Fig. \ref{StackAB} at fixed direction
$\vec{n}$ of light incidence. In Fig. \ref{DRn_YV} we chose%
\begin{equation}
n_{x}=n_{y}=n_{0}\text{, \ where }n_{0}=-0.493489, \label{n0}%
\end{equation}
because this particular direction of light incidence produces the axially
frozen mode regime at certain frequency $\omega_{0}$, shown in Fig.
\ref{DRn_YV_f}. Due to the relatively small anisotropy of yttrium vanadate,
the axial dispersion relation in Fig. \ref{DRn_YV} displays rather weak
asymmetry $\omega\left(  k_{z}\right)  \neq\omega\left(  -k_{z}\right)  $,
which makes it virtually impossible to develop a stationary inflection point
at the lowest spectral branches. As we go to upper photonic bands, the
situation improves. In Fig. \ref{DRn_YV_f} we present the enlarged fragment of
the axial dispersion relation in Fig. \ref{DRn_YV}. This fragment covers the
boundary region between the forth and the fifth bands. At frequency%
\begin{equation}
\omega_{0}=9.164450223c/L \label{w0}%
\end{equation}
one of the spectral branches develops axial stationary inflection point
(\ref{ASIP}), associated with the possibility of the axial frozen mode regime.

Our numerical analysis based on the transfer matrix approach (the
computational details are presented in the next section) indeed shows a very
robust (axial) frozen mode regime in this setting, in spite of the fact that
the dielectric anisotropy of yttrium vanadate is more than an order of
magnitude short of the optimal value. The drawback though is that the
frequency bandwidth of the effect is roughly an order of magnitude narrower,
compared to what could be achieved with hypothetical materials displaying much
stronger anisotropy at optical frequencies.

Let us start with the results presented in Figs. \ref{tr_AM_TE} and
\ref{tr_AM_TH}. The bottom plots in both figures display the frequency
dependence of the stack transmittance $\tau$ for two different polarizations
of incident light . Clearly, in the vicinity of the frozen mode frequency
$\omega_{0}$, the transmittance remains significant, which implies that a
significant portion of the incident radiation is converted into the frozen
mode. The top plots in Figs. \ref{tr_AM_TE} and \ref{tr_AM_TH} display the
amplitude of the two Bloch components of the transmitted wave $\Psi_{T}$. The
solid and dotted lines correspond to the propagating and evanescent Bloch
components, respectively. In the vicinity of the frozen mode frequency (at
$\omega_{a}<\omega<\omega_{b}$), there is one propagating component
($\Psi_{pr}$) and one evanescent component ($\Psi_{ev}$), each of which
diverges as $\omega\rightarrow\omega_{0}$, in accordance with Eq.
(\ref{W(w)}). At $\omega<\omega_{a}$ and $\omega>\omega_{b}$, the transmitted
wave $\Psi_{T}$ is a superposition of two propagating components with
different group velocities, which constitutes the phenomenon of double
refraction. The characteristic frequencies in Figs. \ref{tr_AM_TE} and
\ref{tr_AM_TH} are explained in Fig. \ref{DRn_YV_f}.

Fig. \ref{AM0_YV_TE} shows the frequency dependence of the resulting field
amplitude $\left\vert \Psi_{T}\left(  0\right)  \right\vert ^{2}$ at the slab
boundary, along with the amplitudes of its propagating and evanescent
components $\left\vert \Psi_{pr}\left(  0\right)  \right\vert ^{2}$ and
$\left\vert \Psi_{ev}\left(  0\right)  \right\vert ^{2}$. Although each of the
two Bloch contributions to $\Psi_{T}\left(  0\right)  $ diverges as
$\omega\rightarrow\omega_{0}$, their superposition $\Psi_{T}\left(  0\right)
$ remains finite, to meet the boundary conditions (\ref{BC}) at $z=0$. As we
move further away from the slab boundary, the evanescent component dies out,
while the propagating mode $\Psi_{pr}\left(  z\right)  $ remains constant and
large. As a result, the destructive interference of $\Psi_{pr}\left(
z\right)  $ and $\Psi_{ev}\left(  z\right)  $ is removed, and the resulting
field amplitude $\left\vert \Psi_{T}\left(  z\right)  \right\vert ^{2}$ grows
and approaches the value $\left\vert \Psi_{pr}\right\vert ^{2}$. This scenario
is illustrated in Fig. \ref{AMz_YV_TE}\emph{b} and \ref{AMz_YV_TE}\emph{c}. If
the frequency $\omega$ of the incident wave and its direction of propagation
$\vec{n}$ exactly correspond to the critical values $\omega_{0}$ and $\vec
{n}_{0}$, then the electromagnetic field $\Psi_{T}\left(  z\right)  $ inside
the semi-infinite stack is described by a linearly diverging non-Bloch Floquet
eigenmode%
\[
\left\vert \Psi_{T}\right\vert ^{2}\propto z^{2},
\]
as shown in Fig. \ref{AMz_YV_TE}\emph{a}.

By way of example, let us present the actual geometrical parameters of the
stack supporting the axially frozen mode regime for the case of infrared light
with $\lambda=1550$ nm and the direction of incidence (\ref{n0}). The
expression (\ref{w0}) for the frozen mode frequency yield%
\[
d=L/2=\allowbreak1\allowbreak130.4\text{ nm.}%
\]
In practice, we do not have to adjust the layer thicknesses in order to
achieve the frozen mode regime at a prescribed wavelength. Instead, we can
tune the system into the axially frozen mode regime by adjusting the direction
$\vec{n}$ of light incidence.

\textbf{Acknowledgments}. The effort of A. Figotin and I. Vitebskiy was
supported by the U.S. Air Force Office of Scientific Research under the grant
FA9550-04-1-0359. The authors (JB) also wish to thank the Defense Advanced
Research Projects Agency (DARPA) for support under grant N66001-03-1-8900
through SPAWAR.

\section{APPENDIX. Scattering problem for anisotropic semi-infinite stack}

In this section we briefly discuss the standard procedure we use to do the
electrodynamics of stratified media incorporating anisotropic layers. For more
details, see, for example, \cite{Tmatrix,Abdul99,Abdul00,PRE03} and references therein.

\subsection{Time-harmonic Maxwell equations in periodic layered media}

Our consideration is based on time-harmonic Maxwell equations%
\begin{equation}
\nabla\times\mathbf{E}\left(  x,y,z\right)  =i\frac{\omega}{c}\mathbf{B}%
\left(  x,y,z\right)  ,\;\nabla\times\mathbf{H}\left(  x,y,z\right)
=-i\frac{\omega}{c}\mathbf{D}\left(  x,y,z\right)  \label{THME}%
\end{equation}
with linear constitutive relations%
\begin{equation}
\mathbf{D}\left(  x,y,z\right)  =\hat{\varepsilon}\left(  z\right)
\mathbf{E}\left(  x,y,z\right)  ,\ \mathbf{B}\left(  x,y,z\right)  =\hat{\mu
}\left(  z\right)  \mathbf{H}\left(  x,y,z\right)  . \label{MR}%
\end{equation}
In layered media, the tensors $\hat{\varepsilon}$ and $\hat{\mu}$ in Eq.
(\ref{MR}) depend on a single Cartesian coordinate $z$. Plugging Eq.
(\ref{MR}) into (\ref{THME}) yields%
\begin{equation}
\nabla\times\mathbf{E}\left(  x,y,z\right)  =i\frac{\omega}{c}\hat{\mu}\left(
z\right)  \mathbf{H}\left(  x,y,z\right)  ,\;\nabla\times\mathbf{H}\left(
x,y,z\right)  =-i\frac{\omega}{c}\hat{\varepsilon}\left(  z\right)
\mathbf{E}\left(  x,y,z\right)  . \label{MEz}%
\end{equation}
Solutions for Eq. (\ref{MEz}) are sought in the following form%
\begin{equation}
\mathbf{E}\left(  x,y,z\right)  =e^{i\left(  k_{x}x+k_{y}y\right)  }\vec
{E}\left(  z\right)  ,\ \mathbf{H}\left(  x,y,z\right)  =e^{i\left(
k_{x}x+k_{y}y\right)  }\vec{H}\left(  z\right)  , \label{LEM}%
\end{equation}
$\allowbreak$which is a standard choice for the scattering problem of a plane
electromagnetic wave incident on a plane-parallel stratified slab. Indeed, in
such a case, due to the boundary conditions (\ref{BC1}), the tangential
components $\left(  k_{x},k_{y}\right)  $ of the wave vector are the same for
the incident, reflected and transmitted waves. The substitution (\ref{LEM}) in
Eq. (\ref{MEz}) allows to separate the tangential field components
$E_{x},E_{y},H_{x},H_{y}$ into a closed system of four linear differential
equations%
\begin{equation}
\partial_{z}\Psi\left(  z\right)  =i\frac{\omega}{c}M\left(  z\right)
\Psi\left(  z\right)  ,\text{ \ where }\;\Psi\left(  z\right)  =\left[
\begin{array}
[c]{c}%
E_{x}\left(  z\right) \\
E_{y}\left(  z\right) \\
H_{x}\left(  z\right) \\
H_{y}\left(  z\right)
\end{array}
\right]  \label{ME4}%
\end{equation}
The $4\times4$ matrix $M\left(  z\right)  $ is referred to as the Maxwell
operator. The reduced Maxwell equation (\ref{ME4}) for the four tangential
field components $\Psi\left(  z\right)  $ should be complemented with the
following expressions for the normal components of the fields%
\begin{equation}%
\begin{array}
[c]{c}%
E_{z}=\left(  -n_{x}H_{y}+n_{y}H_{x}-\varepsilon_{xz}^{\ast}E_{x}%
-\varepsilon_{yz}^{\ast}E_{y}\right)  \varepsilon_{zz}^{-1},\\
H_{z}=\left(  n_{x}E_{y}-n_{y}E_{x}-\mu_{xz}^{\ast}H_{x}-\mu_{yz}^{\ast}%
H_{y}\right)  \mu_{zz}^{-1}.
\end{array}
\label{EzHz}%
\end{equation}
where%
\begin{equation}
n_{x}=ck_{x}/\omega,\ \ \ n_{y}=ck_{y}/\omega. \label{nx, ny}%
\end{equation}

The expression for the Maxwell operator $M\left(  z\right)  $ is very
cumbersome, its explicit form can be found in \cite{PRE03}. The matrix
elements of $M\left(  z\right)  $ depend on the following parameters: the
frequency $\omega$, the direction $\vec{n}$ of light incidence, the material
tensors $\hat{\varepsilon}\left(  z\right)  $ and $\hat{\mu}\left(  z\right)
$.

In a periodic layered medium%
\begin{equation}
M\left(  z+L\right)  =M\left(  z\right)  ,\label{A(z+L)}%
\end{equation}
where $L$ is the stack period. For any given $k_{x}$, $k_{y}$ and $\omega$,
the system (\ref{ME4}) of four ordinary linear differential equations with
periodic coefficients has four Bloch solutions%
\begin{equation}
\Psi_{k_{i}}\left(  z+L\right)  =e^{ik_{i}L}\Psi_{k_{i}}\left(  z\right)
,\ \ i=1,2,3,4\label{Bloch}%
\end{equation}
where $k_{i},~i=1,2,3,4$ correspond to four solutions for $k_{z}$ for given
$k_{x}$, $k_{y}$ and $\omega$%
\begin{equation}
k_{x},k_{y},\omega\longleftrightarrow\left\{  k_{1z},k_{2z},k_{3z}%
,k_{4z}\right\}  =\left\{  k_{1z}^{\ast},k_{2z}^{\ast},k_{3z}^{\ast}%
,k_{4z}^{\ast}\right\}  .\label{w <->  kz}%
\end{equation}
Real $k_{z}$ in (\ref{w <-> kz}) relate to propagating Bloch eigenmodes, while
complex $k_{z}$ relate to the evanescent modes. In the case of propagating
eigenmodes, the correspondence between $k_{z}$ and $\omega$ for fixed
$k_{x},k_{y}$ is referred to as the axial dispersion relation, the concise
form of which is given by Eqs. (\ref{ADR,k}) or (\ref{ADR,n}).

The reduced Maxwell equations (\ref{ME4}) in periodic layered media are
analyzed and solved using the transfer matrix formalism, the detailed
description of which can be found in the extensive literature on the subject
(see, for example, \cite{PRE03,Abdul00,Abdul99,Tmatrix} and references therein).

\subsection{Boundary conditions}

The boundary conditions at the slab/vacuum interface reduce to the continuity
requirement for the tangential field components at $z=0$%
\begin{equation}%
\begin{array}
[c]{c}%
\left(  \mathbf{E}_{I}\left(  x,y,0\right)  \right)  _{\perp}+\left(
\mathbf{E}_{R}\left(  x,y,0\right)  \right)  _{\perp}=\left(  \mathbf{E}%
_{T}\left(  x,y,0\right)  \right)  _{\perp},\\
\left(  \mathbf{H}_{I}\left(  x,y,0\right)  \right)  _{\perp}+\left(
\mathbf{H}_{R}\left(  x,y,0\right)  \right)  _{\perp}=\left(  \mathbf{H}%
_{T}\left(  x,y,0\right)  \right)  _{\perp},
\end{array}
\ \ \label{BC1}%
\end{equation}
where the indices $I$, $R$ and $T$ denote the incident, reflected and
transmitted waves, respectively. Using representation (\ref{LEM}), we can
recast (\ref{BC1}) in a compact form%
\begin{equation}
\Psi_{I}\left(  0\right)  +\Psi_{R}\left(  0\right)  =\Psi_{T}\left(
0\right)  \label{BC}%
\end{equation}
where%
\begin{align}
\Psi_{I}  &  =\left[
\begin{array}
[c]{c}%
E_{I,x}\\
E_{I,y}\\
H_{I,x}\\
H_{I,y}%
\end{array}
\right]  =\left[
\begin{array}
[c]{c}%
E_{I,x}\\
E_{I,y}\\
-E_{I,x}n_{x}n_{y}n_{z}^{-1}-E_{I,y}\left(  1-n_{x}^{2}\right)  n_{z}^{-1}\\
E_{I,x}\left(  1-n_{y}^{2}\right)  n_{z}^{-1}+E_{I,y}n_{x}n_{y}n_{z}^{-1}%
\end{array}
\right]  ,\label{Phi V}\\
\Psi_{R}  &  =\left[
\begin{array}
[c]{c}%
E_{R,x}\\
E_{R,y}\\
H_{R,x}\\
H_{R,y}%
\end{array}
\right]  =\left[
\begin{array}
[c]{c}%
E_{R,x}\\
E_{R,y}\\
E_{R,x}n_{x}n_{y}n_{z}^{-1}+E_{R,y}\left(  1-n_{x}^{2}\right)  n_{z}^{-1}\\
-E_{R,x}\left(  1-n_{y}^{2}\right)  n_{z}^{-1}-E_{R,y}n_{x}n_{y}n_{z}^{-1}%
\end{array}
\right]  .\nonumber
\end{align}
describe the incident and reflected waves, respectively. $n_{x}$, $n_{y}$,
$n_{z}$ are the Cartesian components of the unit vector (\ref{n(k)}).

Knowing the eigenmodes (\ref{Bloch}) inside the slab and using the boundary
conditions (\ref{BC}) we can express the amplitude and composition of the
transmitted wave $\Psi_{T}$ and reflected wave $\Psi_{R}$, in terms of the
amplitude and polarization of the incident wave $\Psi_{I}$. This gives us the
electromagnetic field distribution $\Psi_{T}\left(  z\right)  $ inside the
layered medium, as well as the transmittance and reflectance coefficients of
the semi-infinite slab as functions of the incident wave polarization, the
direction $\vec{n}$ of incidence, and the frequency $\omega$.

\subsection{Energy flux, reflectance, transmittance}

The real-valued Poynting vector is defined by%
\begin{equation}
\vec{S}\left(  x,y,z\right)  =\frac{1}{2}\Re \left[
\mathbf{E}^{\ast}\left(  x,y,z\right)  \times\mathbf{H}\left(  x,y,z\right)
\right]  .\label{Pt}%
\end{equation}
Plugging the representation (\ref{LEM}) for $\mathbf{E}\left(  x,y,z\right)  $
and $\mathbf{H}\left(  x,y,z\right)  $ in Eq. (\ref{Pt}) yields%
\begin{equation}
\vec{S}\left(  x,y,z\right)  =\vec{S}\left(  z\right)  =\frac{1}%
{2}\Re \left[  \vec{E}^{\ast}\left(  z\right)  \times\vec
{H}\left(  z\right)  \right]  ,\label{S=[EH]}%
\end{equation}
implying that none of the three Cartesian components of the energy density
flux $\vec{S}$ depends on the tangential coordinates $x$\ and $y$. In
addition, the energy conservation argument implies that the axial\ component
$S_{z}$ of the energy flux does not depend on the coordinate $z$ either%
\begin{equation}
S_{z}\left(  x,y,z\right)  =S_{z}=\text{const},\;S_{x}\left(  x,y,z\right)
=S_{x}\left(  z\right)  ,\;S_{y}\left(  x,y,z\right)  =S_{y}\left(  z\right)
.\label{S=S(z)}%
\end{equation}
This only apply to the case of a plane monochromatic wave incident on a
lossless layered medium. The explicit expression for the $z$ component of the
energy flux (\ref{S=[EH]}) is%
\begin{equation}
S_{z}=\frac{1}{2}\left[  E_{x}^{\ast}H_{y}-E_{y}^{\ast}H_{x}+E_{x}H_{y}^{\ast
}-E_{y}H_{x}^{\ast}\right]  .\label{Sz(Psi)}%
\end{equation}

Let us turn to the scattering problem for semi-infinite slab. Let $\vec{S}%
_{I}$, $\vec{S}_{R}$ and $\vec{S}_{T}$ be the Poynting vectors of the
incident, reflected and transmitted waves, respectively. The energy
conservation imposes the following relation between the normal components of
these three vectors%
\begin{equation}
\left(  \vec{S}_{T}\right)  _{z}=\left(  \vec{S}_{I}\right)  _{z}+\left(
\vec{S}_{R}\right)  _{z}. \label{ST=SI+SR}%
\end{equation}
Since the stack is presumably composed of lossless materials, the $z$
component of the energy flux is independent of coordinates both inside and
outside the stack. In particular, inside the slab we have
\begin{equation}
\text{at \ }z>0\text{: \ }\left(  \vec{S}_{T}\right)  _{z}=\text{const.}
\label{ST = const}%
\end{equation}
The transmittance $\left(  \tau\right)  $ and the reflectance $\left(
\rho\right)  $ of a lossless semi-infinite slab are defined as%
\begin{equation}
\tau=\frac{\left(  \vec{S}_{T}\right)  _{z}}{\left(  \vec{S}_{I}\right)  _{z}%
},\;\;\rho=-\frac{\left(  \vec{S}_{R}\right)  _{z}}{\left(  \vec{S}%
_{I}\right)  _{z}}=1-\tau, \label{tau, rho}%
\end{equation}

\end{document}